\documentclass[prl,twocolumn,showpacs,preprintnumbers,amsmath,amssymb,floats]{revtex4-1}
\usepackage{graphicx}
\usepackage{amsmath}
\usepackage{epstopdf}
\usepackage{amsbsy}
\usepackage{color}
\usepackage{dcolumn}
\usepackage{bm}

\begin{document}

\title{Impurity states and cooperative magnetic order in Fe-based superconductors}

\author{Maria N. Gastiasoro$^1$, P. J. Hirschfeld$^2$, and Brian M. Andersen$^1$}
\affiliation{$^1$Niels Bohr Institute, University of Copenhagen, Universitetsparken 5, DK-2100 Copenhagen,
Denmark\\
$^2$Department of Physics, University of Florida, Gainesville, Florida 32611, USA}

\date{\today}

\begin{abstract}

We study impurity bound states and impurity-induced order in the superconducting state of LiFeAs within a realistic five-band model based on the band structure and impurity potentials obtained from density functional theory (DFT).  In agreement with recent experiments, we find that  Co impurities are too weak produce sub-gap bound states, whereas stronger impurities like Cu do. We also obtain the bound state spectrum for magnetic impurities, such as Mn, and show how spin-resolved tunnelling may determine the nature of the various defect sites in iron pnictides, a prerequisite for using impurity bound states as a probe of the ground state pairing symmetry. Lastly we show how impurities pin both orbital and magnetic order, providing an explanation for a growing set of experimental evidence for unusual magnetic phases in doped iron pnictides.

\end{abstract}


\maketitle

It is crucial to understand the role of disorder in high-temperature superconductors (SC) because the materials are obtained from chemical doping with substitutional impurity atoms. In addition, through the large advance of scanning tunneling microscopy (STM), local perturbations in the host material act as nano-probes of the underlying quantum state. For the Fe-based superconductors (FeSC), a recent series of experiments have measured the local density of states (LDOS) near various impurity sites.\cite{hoffman11} 
In particular, STM measurements within the SC state have focussed largely on FeSe, LiFeAs, and NaFeAs,\cite{song11,hanaguri12,allen12,yang12,grothe12,hanaguriunpub,rosenthal13} revealing a complex pattern of distinct impurity-induced LDOS modulations including unusual sub-gap bound states, local C$_4$ symmetry breaking, and generation of electronic dimers. At present no theoretical model exists which correctly captures the LDOS structure near these different impurity sites.

Theoretically, both potential and magnetic point-like scatterers can generate in-gap bound states in multi-band $s_\pm$-wave SC. The single-impurity problem has been addressed both within simplified two-band models,\cite{ng09,matsumoto09,bang09,tsai09,zhang09,li09,akbari10} and a five-band approach,\cite{kariyado10} reaching, however, different conclusions about the presence and location of in-gap bound states. Recently, an important source of this discrepancy was shown to be the sensitivity of the low-energy states to the band structure and SC gap shape.\cite{beaird12,pjh} For modelling disorder effects in FeSC, it is therefore crucial to include the correct band structure and minimise the sensitivity of the gap structure by self-consistently calculating the SC gaps arising from this band.

A final important recent development is the observation of a component of SDW order observed by e.g. muon spin rotation ($\mu$SR) experiments\cite{bernhard09,bernhard12,wright12} which ``cooperates" with, rather than competes with SC as is commonly 
assumed.  This component, which exists in an intermediate doping range around optimal doping, is evidently correlated with disorder and
disappears above $T_c$.  This type of disorder-induced magnetism supported by the SC state 
is reminiscent of that observed in underdoped cuprates where a wedge-like extension
of quasi-long range antiferromagnetic (AF) order extends into the SC dome.   Within the theory proposed by the present authors\cite{andersen07,christensen11}, this effect is caused by the coherent superposition of droplets of magnetic order which form around each impurity due to both the presence of residual AF correlations in the SC state and to the formation of bound states near the impurities.\cite{Schmid10}

Here we present a first step towards  realistic theoretical modeling of impurity states in LiFeAs, including magnetic correlations, by fixing both the band and the SC pairing constants from the DFT-acquired band structure of this material. The remaining degrees of freedom are associated with the impurity potential $V_{\rm{imp}}$ and the strength of the electronic correlations. Below, we focus first on the LDOS in the uncorrelated SC ($U\!\!=\!\!J\!\!=\!\!0$) and map out the LDOS around nonmagnetic and magnetic impurities in LiFeAs. Second, when including correlations ($U,J\!\neq0$) we show how impurities can locally induce magnetic order,
 and how STM measurements of the LDOS can be used to confirm our picture.  Finally, we discuss how this effect leads to the ``cooperative" SC-SDW coexistence phase\cite{bernhard12,wright12}.

The starting point of the theoretical analysis is the following five-orbital Hamiltonian
\begin{equation}
 \label{eq:H}
 H=H_{0}+H_{int}+H_{BCS}+H_{imp},
\end{equation}
where $H_0$ constitutes the kinetic part obtained from a tight-binding fit to the DFT band-structure of LiFeAs\cite{footnote1}
\begin{equation}
 \label{eq:H0}
H_{0}=\sum_{\mathbf{ij},\mu\nu,\sigma}t_{\mathbf{ij}}^{\mu\nu}c_{\mathbf{i}\mu\sigma}^{\dagger}c_{\mathbf{j}\nu\sigma}-\mu_0\sum_{\mathbf{i}\mu\sigma}n_{\mathbf{i}\mu.\sigma}.
\end{equation}
Here, the operators $c_{\mathbf{i} \mu\sigma}^{\dagger}$ create an electron at the $i$-th site in orbital $\mu$ and spin $\sigma$, and $\mu_0$ is the chemical potential which is fixed so that the doping $\delta=\langle n \rangle - 6.0 = 0.0$.
The indices $\mu$ and $\nu$ run through 1 to 5 corresponding to the Fe orbitals $d_{3z^2-r^2}$, $d_{yz}$, $d_{xz}$, $d_{xy}$, and $d_{x^2-y^2}$.

The second term in Eq.(\ref{eq:H}) describes the Coulomb interactions restricted to intrasite processes
\begin{align}
 \label{eq:Hint}
 H_{int}&=U\sum_{\mathbf{i},\mu}n_{\mathbf{i}\mu\uparrow}n_{\mathbf{i}\mu\downarrow}+(U'-\frac{J}{2})\sum_{\mathbf{i},\mu<\nu,\sigma\sigma'}n_{\mathbf{i}\mu\sigma}n_{\mathbf{i}\nu\sigma'}\\\nonumber
&\quad-2J\sum_{\mathbf{i},\mu<\nu}\vec{S}_{\mathbf{i}\mu}\cdot\vec{S}_{\mathbf{i}\nu}+J'\sum_{\mathbf{i},\mu<\nu,\sigma}c_{\mathbf{i}\mu\sigma}^{\dagger}c_{\mathbf{i}\mu\bar{\sigma}}^{\dagger}c_{\mathbf{i}\nu\bar{\sigma}}c_{\mathbf{i}\nu\sigma},
\end{align}
which includes the intraorbital (interorbital) interaction $U$ ($U'$), the Hund's rule coupling $J$ and the pair hopping energy $J'$.
We assume orbitally rotation-invariant interactions $U'=U-2J$ and $J'=J$.

\begin{figure}[t]
\begin{center}
\includegraphics[width=7.85cm]{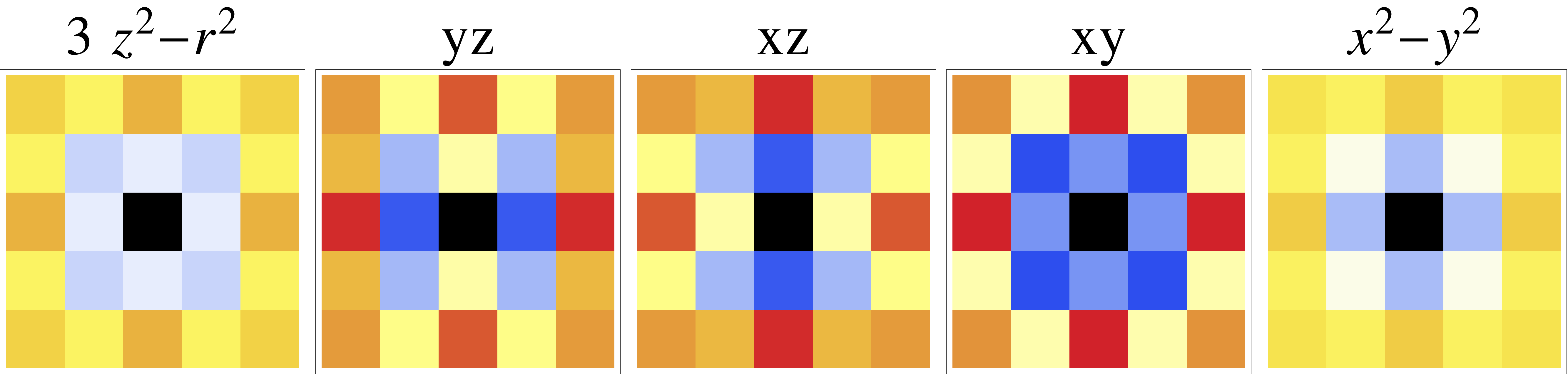}
\includegraphics[width=0.7cm]{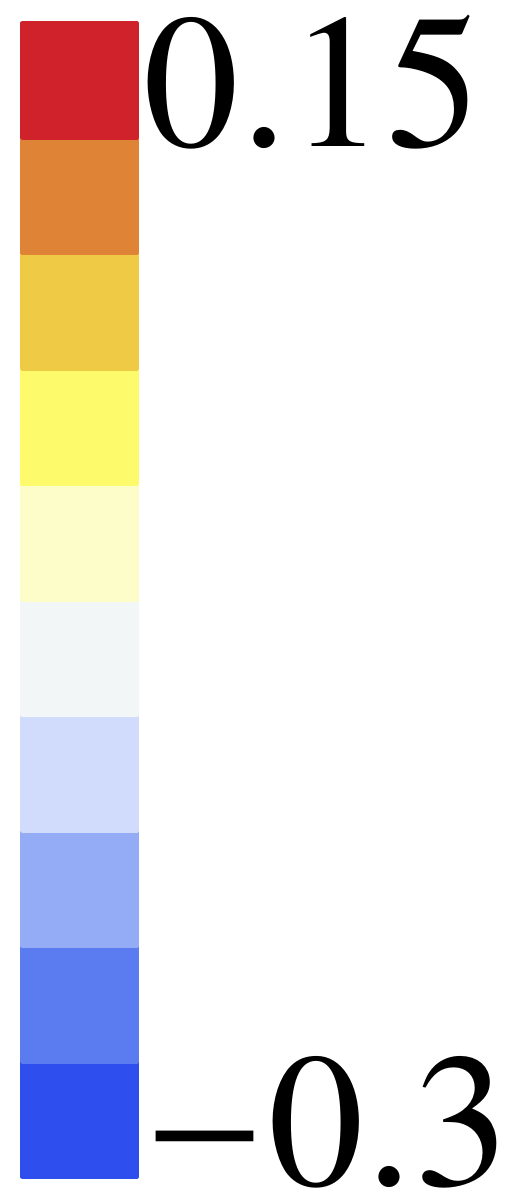}
\\
\includegraphics[width=7.5cm]{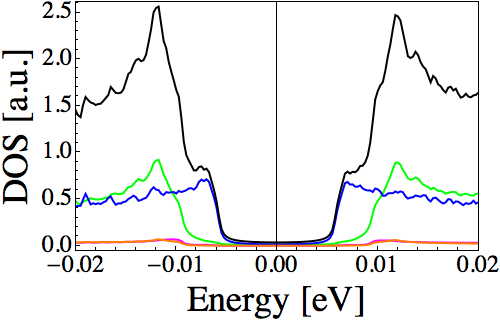}
\end{center}
\caption{(Color online) Top: spatial dependence of the intraorbital effective pairing constants $\Gamma^{\mu\mu}_{\mu\mu}(\mathbf{r_{ij}})$ in eV. The black center site is repulsive with $\sim2.5$eV. Bottom: DOS for the homogeneous SC phase showing the total (black line) and orbitally resolved DOS (green: $d_{yz}/d_{xz}$; blue: $d_{xy}$; orange: $d_{x^2-y^2}$; magenta: $d_{z^2}$).}
\label{fig:1}
\end{figure}

The third term in Eq.(\ref{eq:H}) is given by
\begin{equation}
 H_{BCS}=-\sum_{\mathbf{i}\neq \mathbf{j},\mu\nu}[\Delta_{\mathbf{ij}}^{\mu\nu}c_{\mathbf{i}\mu\uparrow}^{\dagger}c_{\mathbf{j}\nu\downarrow}^{\dagger}+\mbox{H.c.}],
\end{equation}
with SC order parameter $\Delta_{\mathbf{ij}}^{\mu\nu}=\sum_{\alpha\beta}\Gamma_{\mu\alpha}^{\beta\nu}(\mathbf{r_{ij}})\langle\hat{c}_{\mathbf{j}\beta\downarrow}\hat{c}_{\mathbf{i}\alpha\uparrow}\rangle$. Here $\Gamma_{\mu\alpha}^{\beta\nu}(\mathbf{r_{ij}})$ denotes the effective pairing strength between sites (orbitals) $\mathbf{i}$ and $\mathbf{j}$ ($\mu$, $\nu$, $\alpha$ and $\beta$) obtained from the RPA spin- $\chi^{RPA}_s$ and charge susceptibilities $\chi^{RPA}_c$ relevant for LiFeAs
\begin{align}
\Gamma_{\mu\alpha}^{\beta\nu}({\mathbf{k}}&-{\mathbf{k}}')=\left[ \frac{3}{2} U^s \chi^{RPA}_s({\mathbf{k}}-{\mathbf{k}}') U^s +  \frac{1}{2} U^s\right.\nonumber\\
&\left.-\frac{1}{2} U^c \chi^{RPA}_c({\mathbf{k}}-{\mathbf{k}}') U^c +  \frac{1}{2} U^c \right]_{\mu\alpha}^{\beta\nu},
\end{align}
where $U^s$ and $U^c$ are $5\times5$ matrices identical to those of Ref.~\onlinecite{graser09}. The real-space pairings are then obtained by $\Gamma_{\mu\alpha}^{\beta\nu}(\mathbf{r_{ij}})=\sum_{\mathbf{q}} \Gamma_{\mu\alpha}^{\beta\nu}({\mathbf{q}}) \exp(i{\mathbf{q}}\cdot({\mathbf{r_i}}-{\mathbf{r_j}}))$ where we retain all possible orbital combinations up to next-nearest neighbors (NNN). For the present band, the RPA susceptibilities are strongly peaked near $(0,\pm\pi)$ and $(\pm\pi,0)$ favoring an $s^\pm$ pairing state. In Fig.~\ref{fig:1}(a) we show the spatial dependence of the dominant intraorbital pairings $\Gamma_{\mu\mu}^{\mu\mu}(\mathbf{r_{ij}})$ obtained when $J=U/4$ and $U=0.865$eV which yield a fully gapped $s^\pm$ phase with a two-gap peak-structure as seen from the total DOS in Fig.~\ref{fig:1}(b). It is striking that a very similar DOS has been recently measured in LiFeAs by several groups.\cite{hanaguri12,grothe12,allen12} As seen from Fig.~\ref{fig:1}(b) the inner coherence peaks are dominated by the $d_{xy}$ orbital, whereas the outer large-gap coherence peaks consist of significant contributions from both the $d_{xy}$ and $d_{xz}$/$d_{yz}$ orbitals. In momentum space, the peaks at lower energy arise from a smaller gap on the outermost hole pocket around $\Gamma$, which is mainly $d_{xy}$, as opposed to a larger gap on the inner hole pockets around $\Gamma$ and the electron pocket around $M$, which consist primarily of $d_{xy}$ and $d_{xz}$/$d_{yz}$ weight. This agrees with recent ARPES measurements,\cite{borisenko,umezawa12} STM quasi-particle interference (QPI),\cite{allen12} and other theoretical studies.\cite{yan13}

The last term in the Hamiltonian (\ref{eq:H}) describes the effect of a point-like impurity
\begin{equation}
 H_{imp}=\sum_{\mu\sigma} V_{\rm{imp}}^{\mu\sigma} c_{\mathbf{i^*}\mu\sigma}^{\dagger}c_{\mathbf{i^*}\mu\sigma}\label{Himp},
\end{equation}
which adds a local potential $V_{\rm{imp}}^{\mu\sigma}$ at a site $\mathbf{i^*}$ on orbital $\mu$ with spin $\sigma$. We include only intraorbital terms in Eq.(\ref{Himp}) consistent with first principles studies of transition metal atoms in LaFeAsO\cite{nakamura} and LiFeAs.\cite{tom}

In the absence of correlations, the dependence of the LDOS on $V_{\rm{imp}}$ can be most easily obtained within the so-called T-matrix approach. Here, based on $H=H_{0}+H_{BCS}$ one obtains the free retarded $10\times 10$ Nambu Greens function $G^R_0(k,\omega)=[(\omega+i\eta)I_{10\times 10} - H_{Nambu}]^{-1}$ where $H_{Nambu}=\left(\{H_0(k),H_{BCS}(k)\},\{H^\dagger_{BCS}(k),-H^T_0(-k)\}\right)$. The single impurity problem is solved exactly by the full Greens function given by $G^R(i,j,\omega) = G^R_0(0,\omega) + G^R_0(i,\omega) T(\omega) G^R_0(-j,\omega)$, where $i,j$ denote sites in the lattice and $T(\omega)$ in the T-matrix.

\begin{figure}[t]
\begin{center}
\includegraphics[width=2.81cm]{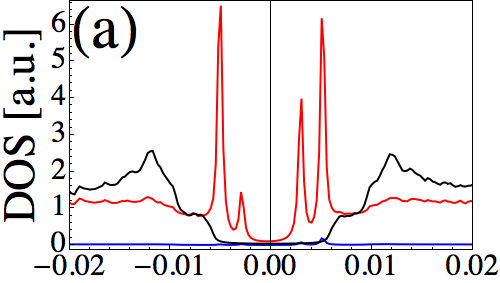}
\includegraphics[width=2.81cm]{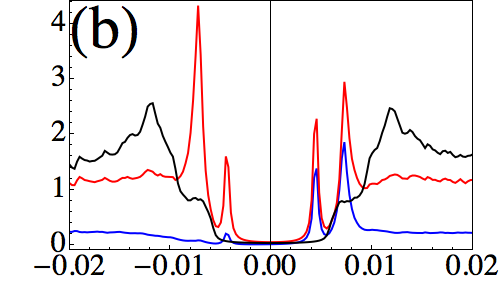}
\includegraphics[width=2.81cm]{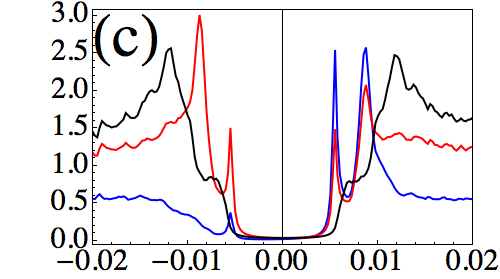}
\\
\includegraphics[width=2.81cm]{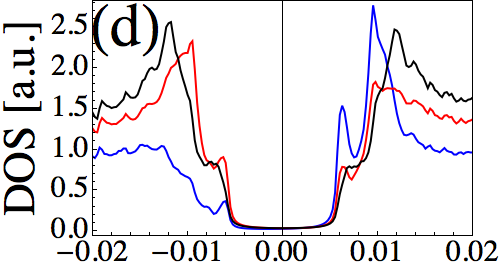}
\includegraphics[width=2.81cm]{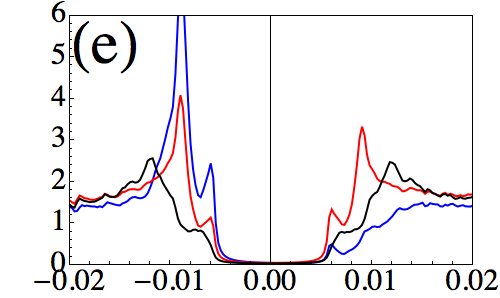}
\includegraphics[width=2.81cm]{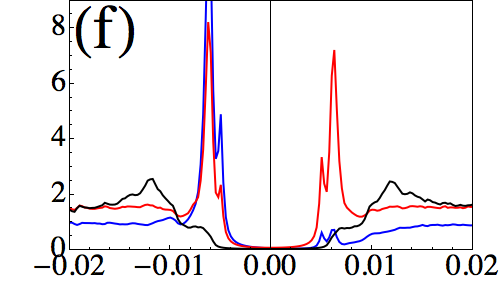}
\\
\includegraphics[width=2.81cm]{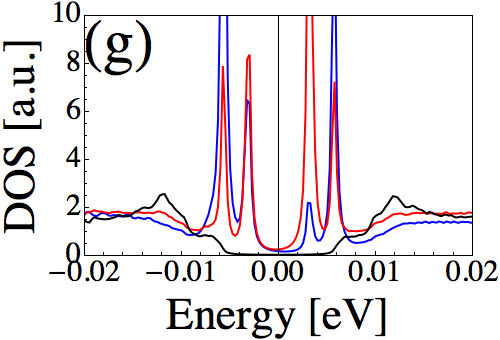}
\includegraphics[width=2.81cm]{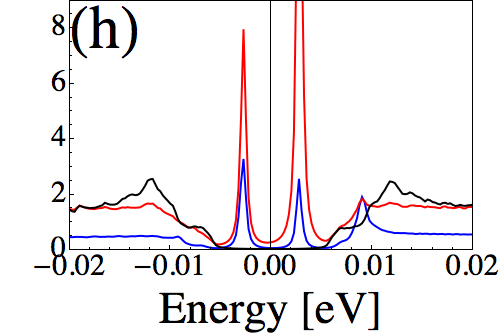}
\includegraphics[width=2.81cm]{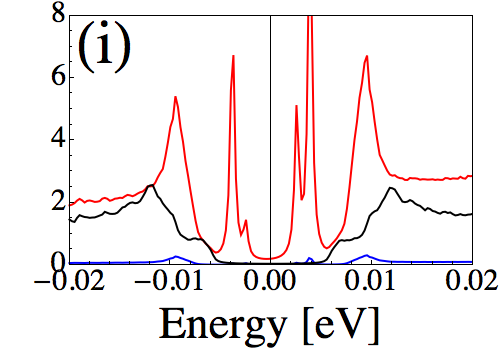}
\\
\end{center}
\caption{(Color online) Total LDOS for nonmagnetic point-like scatterers at the impurity site (blue line) and at the NN site (red line). Panels (a-i) correspond to $V_{\rm{imp}}=-8.0, -1.5,-0.75, -0.4, 0.4, 0.75, 1.5, 2.0, 8.0$eV. In all plots the solid black line is the DOS of the clean SC.}
\label{fig:2}
\end{figure}

Figure~\ref{fig:2} shows the LDOS at the impurity and nearest neighbor (NN) sites for different nonmagnetic scattering strengths $V_{\rm{imp}}$ assumed to be orbitally independent for simplicity.
As seen, in-gap bound states exist for all $|V_{\rm{imp}}|\gtrsim1$eV whereas weaker potentials ($|V_{\rm{imp}}|\lesssim1$eV) mainly cause spectral weight shifts  between the coherence peaks.
Recent STM studies of Co and Cu impurities in superconducting Na(Fe$_{0.97-x}$Co$_{0.03}$Cu$_{x}$)As found distinct LDOS modulations near these Fe substituents.\cite{yang12,yang2013} Very weak spatial variation was reported around Co atoms. Using the effective impurity potential of Co obtained from ${\it ab~ initio}$ calculations ($V_{\rm{imp}}^{Co}\sim$$-0.4$eV)\cite{nakamura,tom}, we find that indeed the LDOS modulation are very weak as seen from Fig.~\ref{fig:2}(d). Near the Cu atoms, the STM study found weak in-gap quasiparticle excitations near the positive gap edge and a suppression of LDOS near the gap edge at negative bias.\cite{yang2013} Within our modelling, and in overall agreement with DFT, this implies that Cu behave as intermediate attractive scatterers, since the resulting LDOS shown in panels Fig.~\ref{fig:2}(b,c) agree with this finding.

\begin{figure}[t]
\begin{center}
\includegraphics[width=2.81cm]{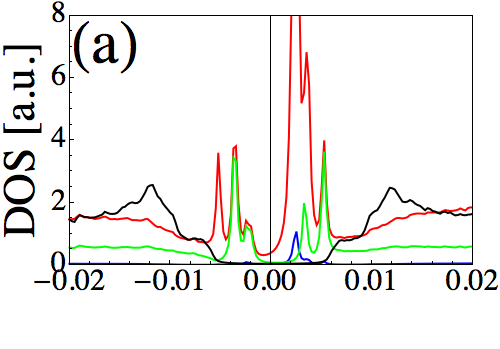}
\includegraphics[width=2.81cm]{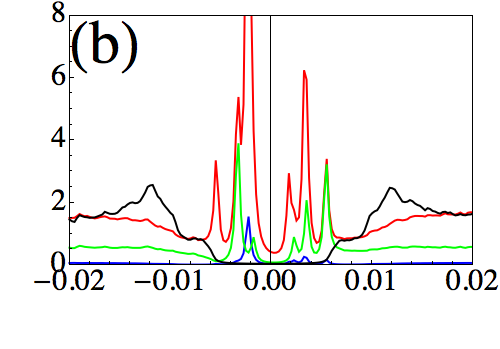}
\includegraphics[width=2.81cm]{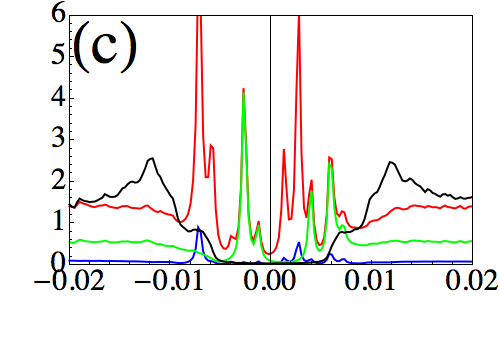}
\\
\includegraphics[width=2.81cm]{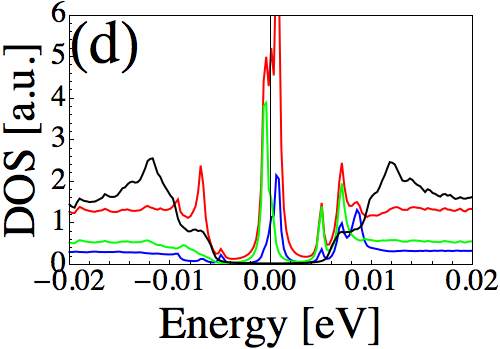}
\includegraphics[width=2.81cm]{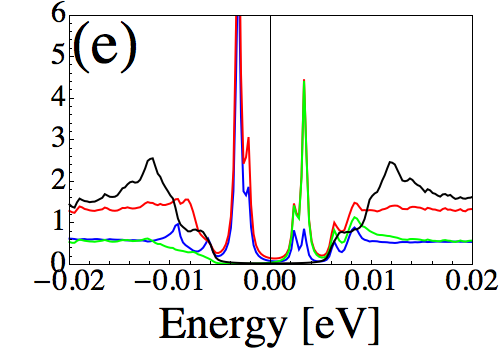}
\includegraphics[width=2.81cm]{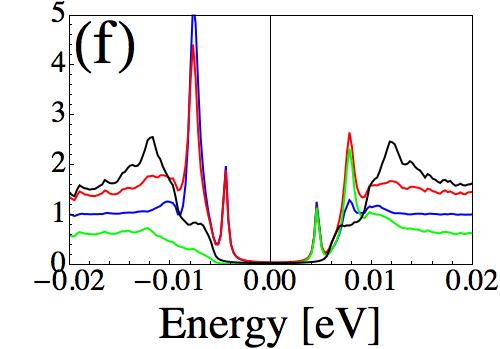}
\end{center}
\caption{(Color online) Total LDOS for magnetic point-like scatterers at the impurity site (blue lines) and at the NN site (red lines). The green lines show the spin-up LDOS at the NN site. Panels (a-f) correspond to $V_{\rm{imp}}=-8.0, -6.0, -4.0, -2.0, -1.0, -0.5$eV. Repulsive potentials lead to the same total LDOS (with interchanged spin-polarisation of the individual bound states).}
\label{fig:3}
\end{figure}

Hanaguri {\it et al.} have located at least six different defect sites on the surface of LiFeAs, some of which induce genuine sub-gap bound states.\cite{hanaguriunpub} As evident from Fig.~\ref{fig:2} these may be caused by intermediate-strong scatterers. However, they can also arise from magnetic impurities.
An STM study on NaFeAs found Curie-like free moment behavior in the case of Mn impurities indicating their magnetic nature.\cite{yang2013} For a single-site magnetic impurity, we show in Fig.~\ref{fig:3} the evolution of the LDOS as a function of the strength of the magnetic scattering potential. In this case, at least four sub-gap bound states are  present for all sizeable impurity potentials. From the panels in Fig.~\ref{fig:3}, it is evident from comparison of the red and green curves, that the LDOS exhibits a striking dependence on the spin-polarisation which may be utilised in future spin-tip polarised STM measurements to unambiguously determine the nature of the scatterers. For example, the absence of any qualitative difference between the measured sub-gap bound states with and without a spin-polarised tip would prove the nonmagnetic nature of the scatterer, and also provide a "smoking gun" for s$\pm$-wave pairing symmetry in FeSC.

We turn now to the study of disorder in the presence of electronic correlations. A mean-field decoupling of Eq.\eqref{eq:Hint} leads to the following multi-band Bogoliubov de-Gennes (BdG) equations\cite{gastiasoro13}

\begin{align}
\sum_{\mathbf{j}\nu}
\begin{pmatrix}
H^{\mu\nu}_{\mathbf{i} \mathbf{j} \sigma} & \Delta^{\mu\nu}_{\mathbf{i} \mathbf{j}}\\
\Delta^{\mu\nu*}_{\mathbf{i} \mathbf{j}} & -H^{\mu\nu*}_{\mathbf{i} \mathbf{j} \bar{\sigma}}
\end{pmatrix}
\begin{pmatrix}
 u_{\mathbf{j}\nu}^{n} \\ v_{\mathbf{j}\nu}^{n}
\end{pmatrix}=E_{n}
\begin{pmatrix}
 u_{\mathbf{i}\mu}^{n} \\ v_{\mathbf{i}\mu}^{n}
\end{pmatrix},
\end{align}
where
\begin{align}
 H^{\mu\nu}_{\mathbf{i} \mathbf{j} \sigma}&=t_{\mathbf{ij}}^{\mu\nu}+\delta_{\mathbf{ij}}\delta_{\mu\nu}[-\mu_0+\delta_{\mathbf{ii^*}}V_{\rm{imp}}+U \langle n_{\mathbf{i}\mu\bar{\sigma}}\rangle\\\nonumber
\quad&+\sum_{\mu' \neq \mu}(U'\langle n_{\mathbf{i}\mu' \bar{\sigma}}\rangle+(U'-J)\langle n_{\mathbf{i}\mu' \sigma}\rangle)].
 \end{align}
The five-band BdG equations are solved on $28\times28$ lattices with stable solutions found through iterations of the following self-consistency equations $\langle n_{\mathbf{i}\mu\uparrow} \rangle=\sum_{n}|u_{\mathbf{i}\mu}^{n}|^{2}f(E_{n})$, $\langle n_{\mathbf{i}\mu\downarrow} \rangle\!=\!\sum_{n}|v_{\mathbf{i}\mu}^{n}|^{2}(1\!-\!f(E_{n}))$, and $\Delta_{\mathbf{ij}}^{\mu\nu}=\sum_{\alpha\beta}\Gamma_{\mu\alpha}^{\beta\nu}(\mathbf{r_{ij}})\sum_{n}u_{\mathbf{i}\alpha}^{n}v_{\mathbf{j}\beta}^{n*}f(E_{n})$, where $\sum_n$ denotes summation over all eigenstates $n$. When calculating the LDOS we use $20\times20$ supercells to acquire spectral resolution of order $\sim 0.5~\rm{meV}$.

\begin{figure}[t]
\begin{center}
\includegraphics[width=8.cm]{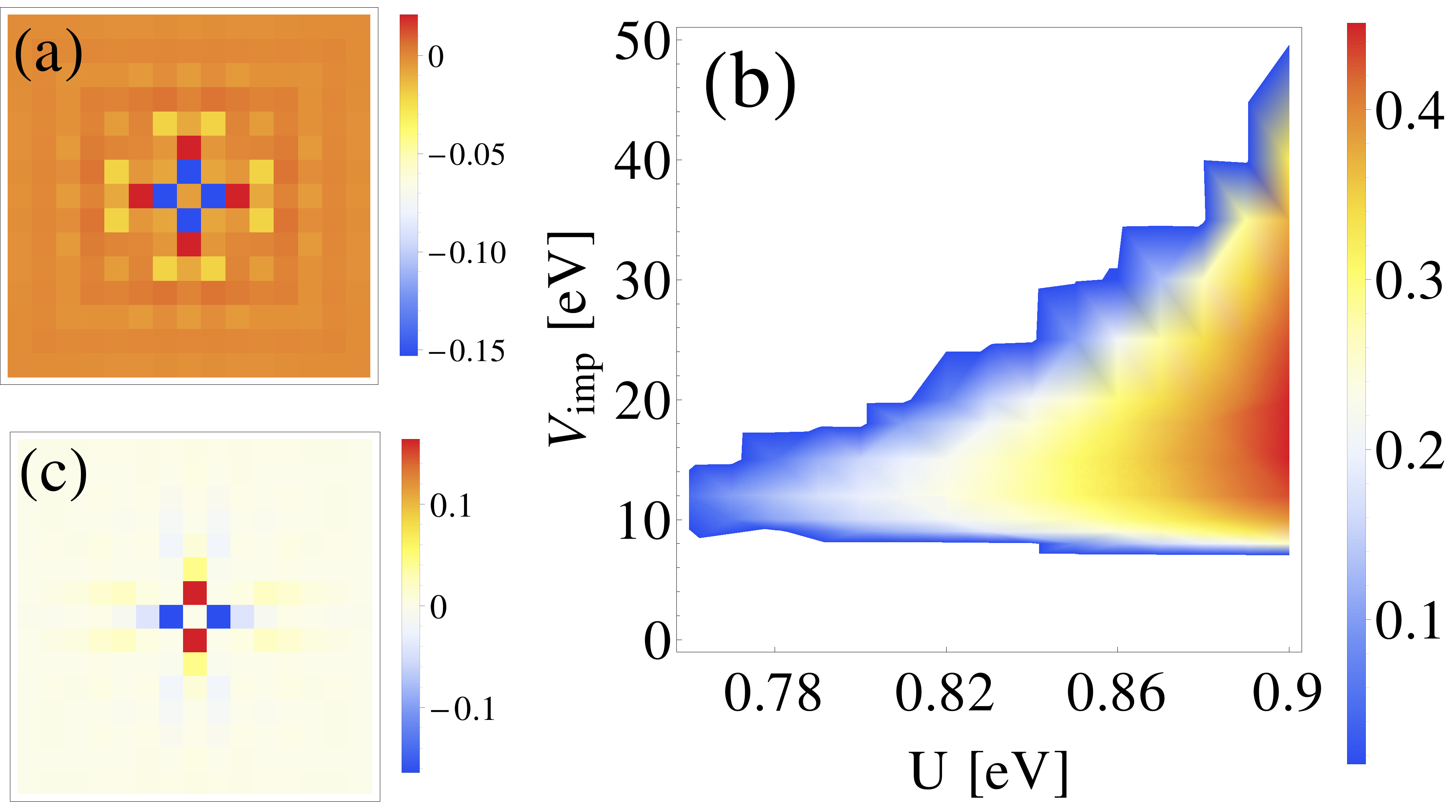}
\\
\vspace{0.2cm}
\includegraphics[width=8.cm]{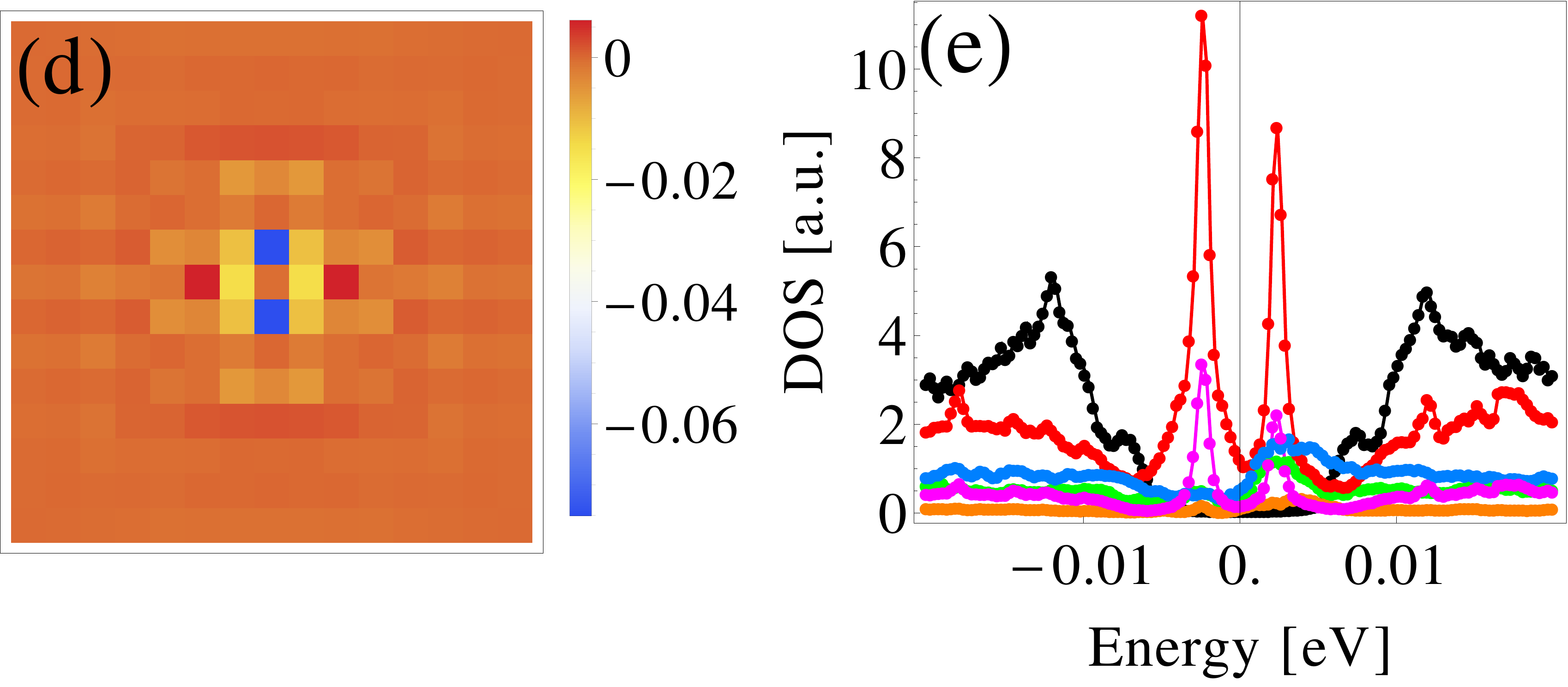}
\includegraphics[width=8.cm]{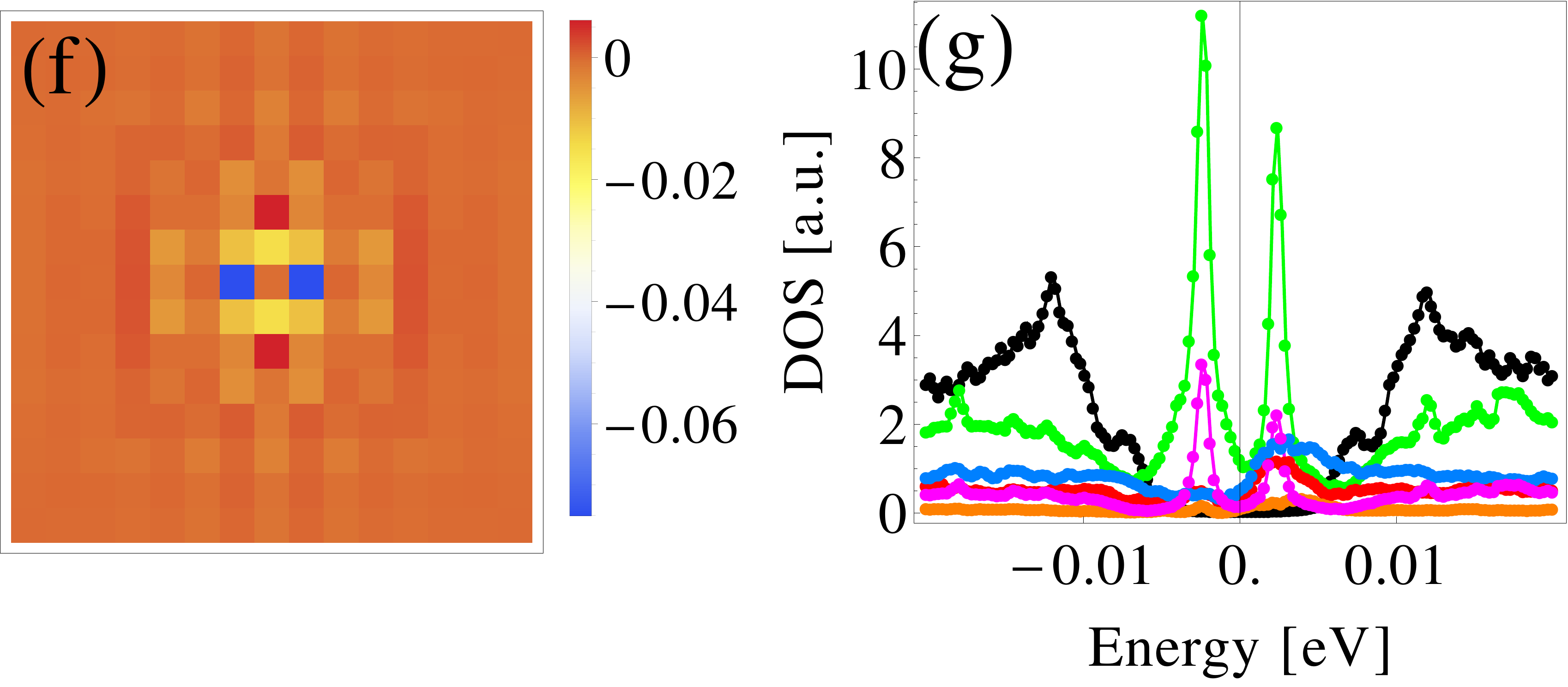}
\end{center}
\caption{(Color online) (a,c) Real-space resolved impurity-induced (a) magnetic order and (c) orbital order ($n_{xz}-n_{yz}$) near a repulsive point-like scatterer with $V_{\rm{imp}}=8.0$eV and  $U=0.865$.
(b) Single impurity phase diagram displaying the impurity induced magnetization at the NN site ($|m_{i^*+1}|$) versus $U$ and $V_{\rm{imp}}$.
(d) Magnetization versus lattice site for the $d_{xz}$ orbital. (e) Orbitally resolved LDOS at the NN sites of the impurity in the $y$ direction. The orbital color code is identical to that in Fig.~\ref{fig:1}(b) [red: $d_{xz}$.] (f) Same as (d) for $d_{yz}$. (g) Same as (e) but at NN site in the $x$ direction.}
\label{fig:4}
\end{figure}

When $U$ is nonzero, but still smaller than the critical value $U_{c2}$ to enter a bulk SDW phase, it mainly acts to suppress the charge modulations.\cite{andersen08}
However, for $U_{c1}\!<\!U\!<\!U_{c2}$ magnetic order may be induced in the vicinity of the impurity as shown in Fig.~\ref{fig:4}(a). The formation of local moments near nonmagnetic scatterers in correlated hosts has been extensively discussed for cuprate SC.\cite{andersen07,Schmid10,tsuchiura01,zhu02,chen04,kontani06,harter07} As seen from the single-impurity phase diagram in Fig.~\ref{fig:4}(b), the impurity-induced magnetization exists only in a finite wedge-shaped region in the $U\!-\!V_{\rm{imp}}$ phase space. The origin of the induced magnetization can be understood from a local crossing of the Stoner instability, and the wedge-shaped region in Fig.~\ref{fig:4}(b) simply reflects the area where the LDOS enhancement at the NN sites is large enough to cross the magnetic instability. By contrast, attractive potentials are unable to support induced magnetization because the LDOS enhancements are too weak. An orbitally resolved analysis reveals that the instability takes place in the $d_{xz}$/$d_{yz}$ orbitals due to large LDOS enhancements at the NN sites for these orbitals as shown in Fig.~\ref{fig:4}(e,g). As a result, the total magnetization is dominated by these two orbitals as seen explicitly from Fig.~\ref{fig:4}(d,f). 

The impurity-induced order found here provides a candidate for the distinct magnetic phases found recently in SC Co-doped BaFe$_2$As$_2$\cite{bernhard09,bernhard12} and NaFeAs\cite{wright12}, and non-SC Mn-doped BaFe$_2$As$_2$.\cite{tucker12,inosov13} For example near optimally doped Ba(Fe$_{1-x}$Co$_x$)$_2$As$_2$, $\mu$SR discovered a disordered inhomogeneous magnetic phase which was not observable by neutrons.\cite{bernhard12} Within the present theoretical scenario, such a phase could be stabilised by a multiple-Co dopant effect similar to the single-impurity result shown in Fig.~\ref{fig:4} where local dopant clusters induce magnetic order but are, however, too weakly coupled to neighboring clusters to be seen by neutron scattering. Stronger coupling between local magnetic regions seem to take place in the Mn-doped samples where neutron scattering measurements have directly detected impurity-induced magnetic ($\pi,\pi$) order.\cite{tucker12}

\begin{figure}[b]
\begin{center}
\includegraphics[width=7.cm]{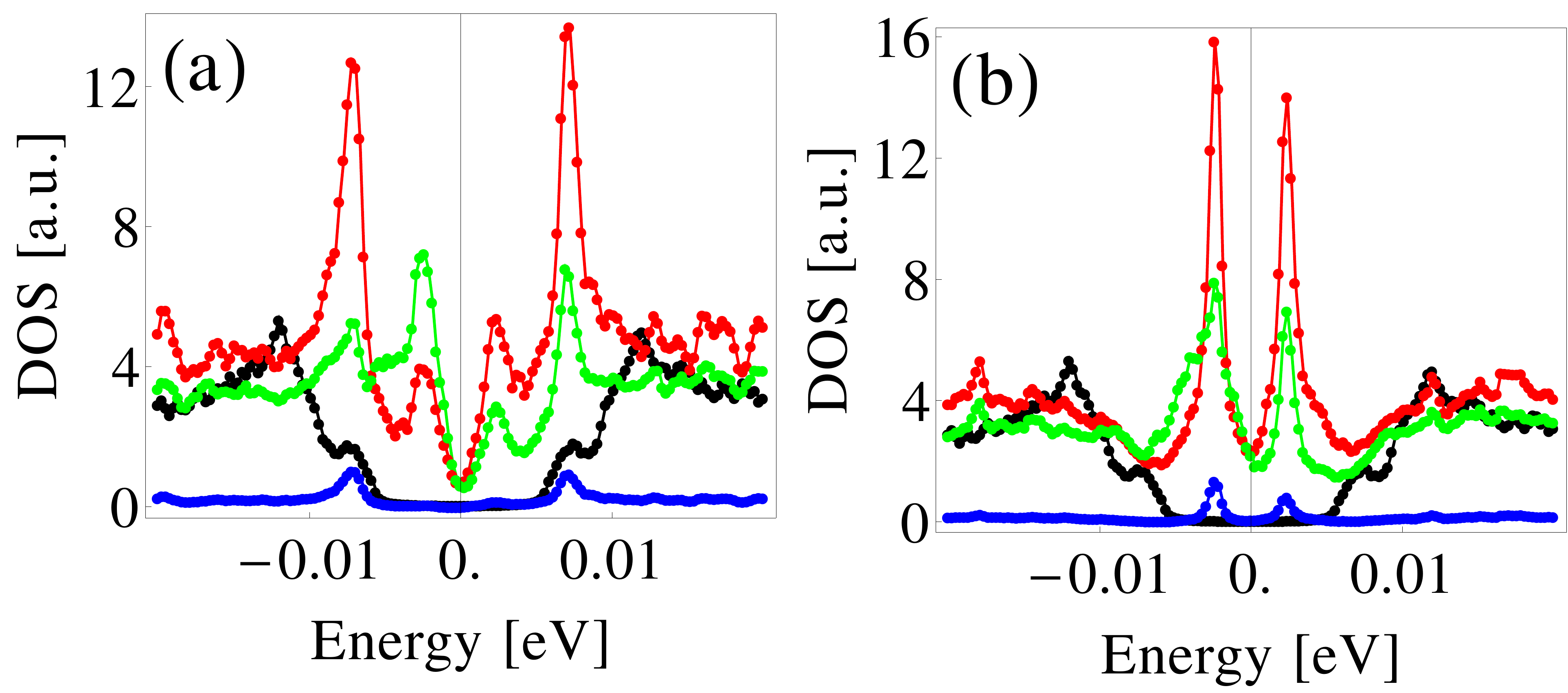}
\includegraphics[width=7.cm]{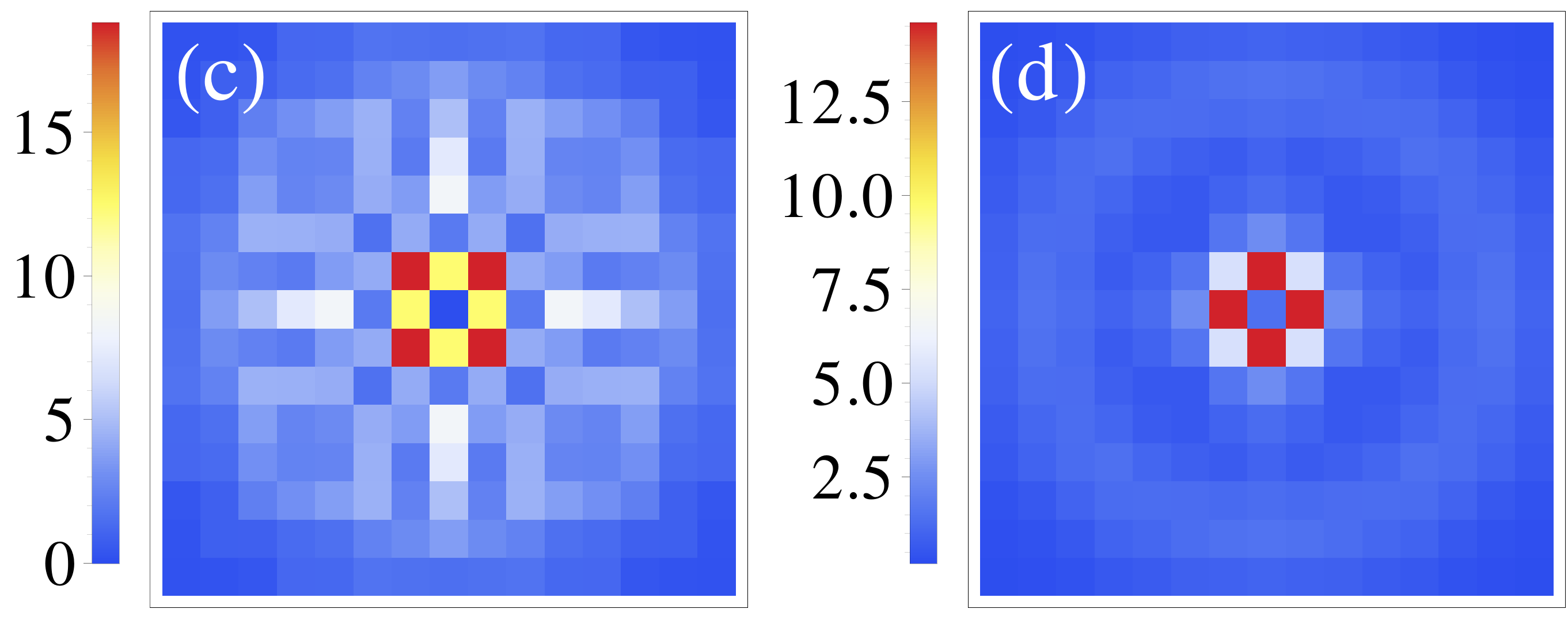}
\end{center}
\caption{(Color online) Comparison of the LDOS near a nonmagnetic scatterer with $V_{\rm{imp}}=8.0$eV in the absence (a,c) and presence (b,d) of induced magnetic order. (a) BdG LDOS versus energy on the impurity site (blue line), at the NN site (red line) and the NNN site (green line) for $U\!=\!0.84\mbox{eV}\!<\!U_{c1}$, (b) Same as (a) for $U\!=\!0.865\mbox{eV}\!>\!U_{c1}$. (c,d) Real-space LDOS maps at $\omega\sim -3$meV for the (a,b) cases.}
\label{fig:5}
\end{figure}

Lastly we return to the topic of bound states and LDOS modulations near disorder sites. The self-consistency and possibility of induced order should alter the T-matrix results presented above. Specifically, the difference between the self-consistent BdG and the (non-selfconsistent) T-matrix approach is the correct spatial profile of $\Delta_{\mathbf{i} \mathbf{j}}$ and the electron density $n_{\mathbf{i}}$ near the impurity within BdG. In addition, only BdG captures the impurity-induced local orbital order ($n_{xz}\neq n_{yz}$) at neighboring sites as shown in Fig.~\ref{fig:4}(c). 

In Fig.~\ref{fig:5}(a) we show the LDOS obtained within self-consistent BdG for the same parameters as in Fig.~\ref{fig:2}(i) and with a sub-critical $U=0.84<U_{c1}$. Compared to Fig.~\ref{fig:2}(i) we note a striking similarity to the non-selfconsistent T-matrix LDOS, which we find to be a general property for all impurity potentials. This ceases to be true, however, when $U>U_{c1}$ causing local magnetic order. As seen from Fig.~\ref{fig:5}(b), the LDOS in the case of induced order pushes essentially all the weight of the outer coherence peaks onto the impurity bound states. This effect is also reflected in the real-space LDOS maps of the bound state wave function shown in Fig.~\ref{fig:5}(c,d).
Without induced order its spectral weight undergoes a $\pi/4$ rotation (from NNN to NN or opposite) under $\omega \rightarrow -\omega$ (see Fig.~\ref{fig:5}(a)). In the presence of induced order however, most of the spectral weight remains at the magnetic NN sites for all bound states.

In summary, we have studied disorder-induced order in FeSC, and provided a first step to realistic modelling of impurity bound states in these materials.
Future studies which combines the DFT-obtained local Wannier states near impurity sites and the present BdG real-space approach constitute a natural next step in the realistic modeling of disorder in FeSC, and hopefully provide a quantitative description of the diverse real-space structures currently observed by STM.

We acknowledge useful discussions with T. Berlijn, T. Hanaguri, and J. Hoffman. B.M.A. and M.N.G. were supported by the Lundbeckfond fellowship (grant A9318). P.J.H. was supported by NSF-DMR-1005625.


\begin{thebibliography}{00}
\bibitem{footnote1} The 3D nature of LiFeAs gives a sizeable $k_z$ dependence of the bands. For the results presented in this paper we have used $k_z=0$.
\end{thebibliography}
\end{document}